\documentclass[%
 reprint,
superscriptaddress,
 amsmath,amssymb,
 aps,
floatfix,
]{revtex4-2}

\usepackage{graphicx}
\usepackage{dcolumn}
\usepackage{bm}
\usepackage{chemformula}
\usepackage{siunitx}

\begin{document}

\title{CVD bilayer graphene spin valves with 26 \si{\micro\metre} spin diffusion length at room temperature}

\author{Timo Bisswanger}
 \email{timo.bisswanger@rwth-aachen.de}
\author{Zachary Winter}
\author{Anne Schmidt}
\author{Frank Volmer}
 \affiliation{2nd Institute of Physics and JARA-FIT, RWTH Aachen University, 52074 Aachen, Germany}
\author{Kenji Watanabe}
 \affiliation{Research Center for Functional Materials, 
National Institute for Materials Science, 1-1 Namiki, Tsukuba 305-0044, Japan}
\author{Takashi Taniguchi}
 \affiliation{International Center for Materials Nanoarchitectonics, National Institute for Materials Science, 1-1 Namiki, Tsukuba 305-0044, Japan}
\author{Christoph Stampfer}
 \affiliation{2nd Institute of Physics and JARA-FIT, RWTH Aachen University, 52074 Aachen, Germany}
 \affiliation{Peter Gr\"unberg Institute (PGI-9) Forschungszentrum J\"ulich, 52425 J\"ulich, Germany}
\author{Bernd Beschoten}
 \email{bernd.beschoten@physik.rwth-aachen.de}
 \affiliation{2nd Institute of Physics and JARA-FIT, RWTH Aachen University, 52074 Aachen, Germany}

\date{\today}

\begin{abstract}
We present inverted spin-valve devices fabricated from CVD-grown bilayer graphene (BLG) that show more than a doubling in device performance at room temperature compared to state-of-the-art bilayer graphene spin-valves. This is made possible by a PDMS droplet-assisted full-dry transfer technique that compensates for previous process drawbacks in device fabrication. Gate-dependent Hanle measurements reveal spin lifetimes of up to \SI{5.8}{\nano\second} and a spin diffusion length of up to \SI{26}{\micro\metre} at room temperature combined with a charge carrier mobility of about \SI{24000}{\square\cm(\volt\second)^{-1}} for the best device. Our results demonstrate that CVD-grown BLG shows equally good room temperature spin transport properties as both CVD-grown single-layer graphene and even exfoliated single-layer graphene.
\end{abstract}

\maketitle

Graphene has proven to be a pivotal material for spintronic devices based on 2D materials. This is largely due to its small intrinsic spin-orbit coupling strength and the high charge carrier mobility, which allow for room temperature spin diffusion lengths over several tens of micrometers.\cite{Han2014Oct,Roche2015Jul,drogeler_12ns_2016} 
Bilayer graphene (BLG) permits additional semiconductor functionality by applying a perpendicular electric displacement field, which creates a gate tunable electronic band gap\cite{Zhang2009Jun} and a spin-orbit gap.\cite{Konschuh2012Mar,Banszerus2020May} These have shown to result in large spin lifetime anisotropies of in-plane and out-of-plane spins,\cite{Xu_strong_2018,Leutenantsmeyer_observation_2018} which can be enhanced when BLG gets proximity-coupled to transition metal dichalcogenides.\cite{Omar2019} 
BLG thus became prominent for new spintronic device concepts including the spin-orbit valve with all-electrical control over spin and orbital degrees of freedom.\cite{Gmitra2017,Ingla-Aynes_SOC-BLG-WSe2_2021}

\begin{figure*}[tb]
\includegraphics{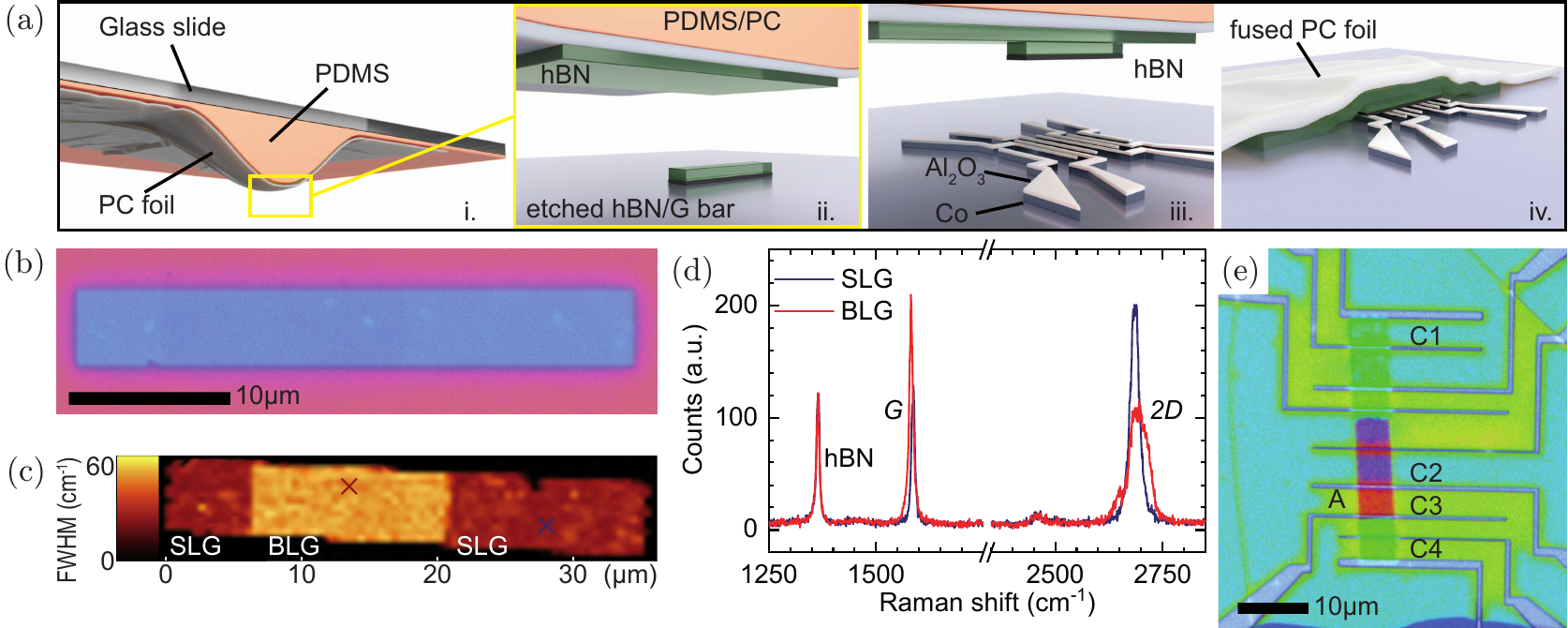}
\caption{(a)~Schematic of the PDMS droplet tool for dry transfer of 2D materials. i.~For the fabrication of inverted spin-vales, a PC-covered PDMS droplet is used to first pick up a large hBN flake. ii.~An etched hBN/CVD graphene (G) bar is picked up. iii.~The whole hBN/hBN/G stack gets transferred onto the prefabricated inverted Co/\ch{Al2O3} contact structure. iv.~The finished device with a fused PC foil on top. (b) Optical image of the etched hBN/CVD graphene stack with (c) a spatial map of the Raman \textit{2D} line full width at half maximum (FWHM) showing BLG at the inner part of the bar. (d) Typical Raman spectra of single and bilayer graphene of the etched hBN/CVD graphene bar taken at the position of the crosses in (c). (e) False color optical image of a Co/\ch{Al2O3}/CVD-BLG/hBN device. Blue parts represent non-suspended BLG parts, meaning that the BLG touches the underlying \ch{SiO2} substrate, whereas red areas highlight suspended BLG parts and green areas SLG. The hBN/G bar is on top of Co/\ch{Al2O3} contacts and underneath a protective hBN and the polymer. C1 to C4 are the electrodes used for transport measurements and A denotes the spin transport region.}
\label{fig:sample}
\end{figure*}

BLG has extensively been used in lateral spin valve devices mainly employing mechanical exfoliation of BLG from natural graphite.\cite{Yang2011,Han_Kawakami_2011,Volmer_role_2013,Volmer_Suppression_2014,drogeler_nanosecond_2014,Volmer_contact-induced_2015,drogeler_nanosecond_2015,Ingla_88percent_2016,Singh_Modulation_2017,Singh_SrO_2017,Leutenantsmeyer_Bias_2018,Leutenantsmeyer_Efficient_2018,Xu_strong_2018,Leutenantsmeyer_observation_2018,Omar2019,Omar2020,Safeer2020}
In comparison, chemical vapor deposition (CVD) is one of the most important growth methods when it comes to integrating synthetically fabricated graphene into scalable spintronic devices.\cite{Backes2020Jan} Recently, there has been significant improvement in the charge and spin transport properties of CVD-grown single-layer graphene (SLG), with device properties on average equally good as those of exfoliated graphene.\cite{banszerus_ultrahigh-mobility_2015,Banszerus2016Feb,drogeler_dry-transferred_2017,Banszerus2019Sep,gebeyehu_spin_2019,KhokhriakovLargeScale2020,Panda_2020} In contrast, until now for BLG this is only valid when considering charge transport properties.\cite{Schmitz2017Jun} The spin transport properties of CVD-grown BLG are still minor compared to both exfoliated BLG and CVD-SLG devices,\cite{Avsar2011,Khokhriakov_Robust_2020,Panda_2020} which is mainly due to technological problems in the growth process as well as in the device fabrication. 

To overcome these limitations we demonstrate an advanced droplet-assisted device fabrication method for non-local inverted spin-valve structures (iSVs). We demonstrate further that the droplet tool enables a reliable dry transfer of pre-patterned hBN/CVD-BLG structures onto prefabricated ferromagnetic Co electrodes with an \ch{Al2O3} barrier. This transfer process is completely free of any chemicals or solvents and is shown to yield excellent charge and spin transport properties in BLG spin valves at room temperature. 

In contrast to conventional lateral spin-valves, where the ferromagnetic electrodes and the oxide barrier are directly evaporated on the 2D materials, the entire contact structure for iSVs is fabricated on \ch{Si^{++}}/\ch{SiO2} wafers beforehand. By placing a hetero-stack of hexagonal boron nitride (hBN) and graphene on the final electrode structure in a dry transfer step, contaminations of the 2D materials from lithographic processes can largely be avoided. This has proven to significantly enhance spin and charge transport parameters in both exfoliated and CVD graphene.\cite{drogeler_nanosecond_2014,drogeler_nanosecond_2015,drogeler_12ns_2016,drogeler_dry-transferred_2017} 

Well-defined bar-shaped graphene is necessary for all lateral spin-valve devices. When solely using mechanical exfoliation this requirement results in the tedious and time-consuming task of optically searching for flakes with an appropriate shape. In contrast, CVD graphene can be grown in flakes whose dimensions are large enough to etch the CVD graphene into the desired shape. This is usually done with reactive ion etching (RIE). The downside of this approach is that the CVD graphene has to be protected during this etching process to maintain its quality. To accomplish this, we recently introduced a two-step process where the CVD graphene first gets delaminated from the copper growth foil by an hBN flake in a dry-transfer step.\cite{drogeler_dry-transferred_2017} 

After being placed onto a Si/\ch{SiO2} wafer, there will be a lithography step to etch the hBN/CVD-graphene into a bar-shaped structure. However, the issue with this patterning process is an observed overall increase in the adhesion of the hBN/CVD-graphene hetero-stack to the underlying substrate, which often prevents a reliable pick-up when using large-area polymer transfer tools. Regarding the observed increased adhesion, we suspect a combination of different effects. On the one hand, the increased temperature in the etching process may lead to stronger adhesion similar to vacuum annealing.\cite{Das-InterfacialBondingCharacteristics-2014,Megra-EnhancementAdhesionEnergy-2021} On the other hand, RIE causes edge functionalization of the 2D materials and chemical modification of the substrate by the ionized gases.\cite{Gongyang_Delamination_2018,Rokni_interfacial-adhesion_2020,Wojtaszek_Hydrogenation_2011} The latter can lead to higher adhesion at the etch boundaries.\cite{Gongyang_Delamination_2018}

We demonstrate that this critical fabrication problem can be solved by using a drop-shaped polydimethylsiloxane (PDMS) stamping tool covered with a poly(bisphenol A carbonate) (PC) film, in the following referred to as the PDMS droplet. Similar droplet tools based on PDMS or other polymers have recently been reported.\cite{Kim2016,Son2018,Frisenda2018,Martanov2020,Wakafuji_2020,Daw2020,Chen_Son_2021} However, they have not been used for the fabrication of spin valves so far. Our PDMS droplet is shown schematically in Fig.\,\ref{fig:sample}a. 

The PDMS droplet tool was made with DowSil Sylgard 184 silicone elastomer. The resin was mixed to a ratio of 8:1 of elastomer base and hardener, and trapped air was removed in a vacuum. Drops of the resin mixture were placed on glass slides using a syringe and hard cured upside down in an oven at \SI{90}{\celsius} for about \SI{30}{\minute}. Polycarbonate (PC, Sigma Aldrich), dissolved in trichloromethane, was spin-coated on glass slides and dried at \SI{90}{\celsius}. The film was peeled off with a hole-punched adhesive tape, stretched over the PDMS droplet, and fixed with the excess adhesive tape (see Fig.\,\ref{fig:sample}a i). PC was chosen because it has a stronger adhesion to 2D materials than other typical transfer polymers such as PMMA, PPC, or PDMS.\cite{wang_one-dimensional_2013,purdie_cleaning_2018,castellanos-gomez_deterministic_2014,pizzocchero_hot_2016} This is important because we need to overcome the aforementioned enhanced adhesion of the hBN/CVD-graphene bar to the Si/\ch{SiO2} substrate. The droplet tool was mounted in a transfer setup. The tool then enables precise and selective control of both the pick-up and deposition of the van der Waals heterostructures because of its small contact area and transparency.

\begin{figure*}[tb]
\includegraphics{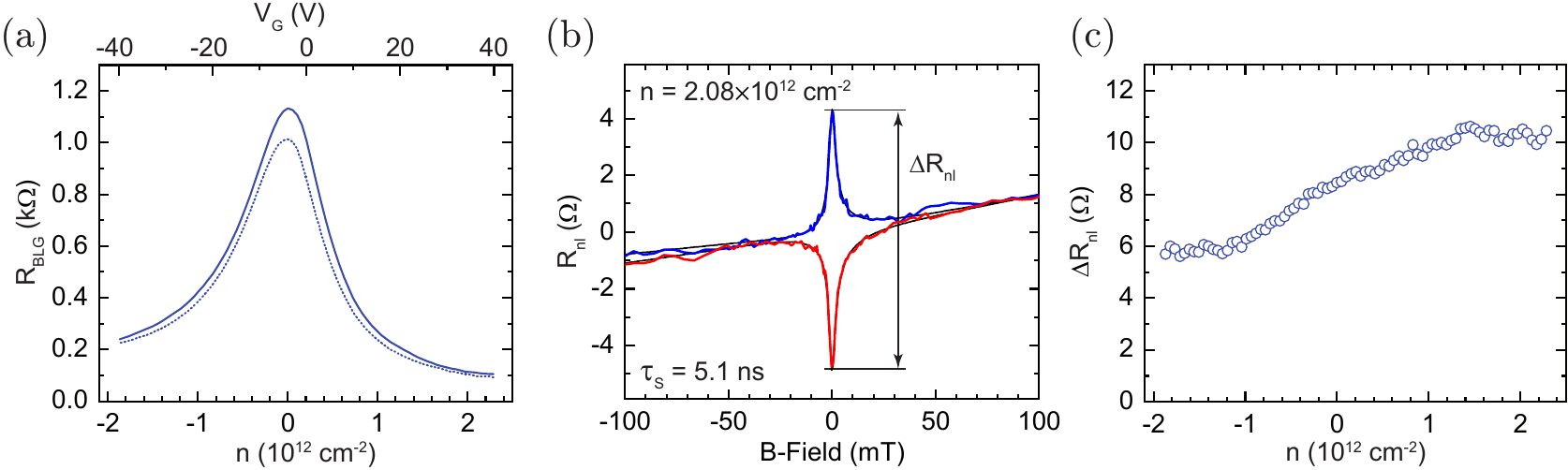}
\caption{(a)~Gate dependent graphene resistance of device 1 (BLG transport region A shown in Fig.\,\ref{fig:sample}e. The trace (dotted line) shows the gate sweep from \SIrange{-40}{40}{\volt}, whereas the retrace (solid line) was swept from \SIrange{40}{-40}{\volt}. (b)~Non-local Hanle spin precession measurement taken at room temperature at $n=\SI{2.08e12}{\per\square\centi\metre}$ ($V_G=\SI{38}{\volt}$). The blue and red curves represent the parallel and antiparallel magnetization configurations of the spin injection and spin detection contacts. The difference at zero magnetic field is known as the non-local spin signal $\Delta R_{nl}$. The thin black curves represent the fitted curve. (c)~Gate dependence of the non-local spin signal $\Delta R_{nl}$.}
\label{fig:measurements}
\end{figure*}

For our study, we use graphene, which was grown by chemical vapor deposition in copper foil enclosures under low pressure\cite{li_large-area_2011,banszerus_ultrahigh-mobility_2015,Schmitz2017Jun,Schmitz2020Sep} with typical sizes of hundreds of micrometers for SLG and tens of micrometers for BLG.\cite{Schmitz2017Jun} The second graphene layer normally starts to grow in the center of a preexisting SLG flake. Therefore, the smaller BLG areas are typically surrounded by larger SLG areas. An etching process can therefore yield devices with sharp junctions between SLG and BLG regions, which will be demonstrated further below. The growth of the CVD graphene is followed by an oxidation step in a humid atmosphere to reduce the adhesion between graphene and the copper surface, which is essential for dry delamination.\cite{banszerus_ultrahigh-mobility_2015,Luo2017May,Schmitz2020Sep} 

The CVD graphene is first transferred onto a clean \ch{Si^{++}}/\ch{SiO2} wafer by a large hBN crystal (thickness $\leq\SI{30}{\nano\metre}$), which has already been exfoliated onto a supporting polymer stack. The stack consists of polydimethylsiloxane (PDMS), polyvinyl alcohol (PVA) and PMMA (for pick-up details cf.\,Refs.\,\citenum{drogeler_dry-transferred_2017} and \citenum{Schmitz2017Jun}). The patterning of the graphene/hBN stack is performed in a RIE process with an \ch{O2}/\ch{CF4} plasma using a PMMA etch mask. Residual cross-linked PMMA is removed by an increased \ch{O2} concentration in the plasma during the final seconds. The PMMA mask is then removed in a warm bath of acetone.

The etched hBN/CVD-graphene bars (see optical image in Fig.\,\ref{fig:sample}b) are characterized by confocal Raman spectroscopy (Figs.\,\ref{fig:sample}c and \ref{fig:sample}d). Single and bilayer graphene regions can reliably be distinguished by the spectral shape of the Raman \textit{2D} line (Fig.\,\ref{fig:sample}d) and the respective \textit{2D} line widths (Fig.\,\ref{fig:sample}c).

In order to transfer these etched hBN/CVD-graphene bars, we first pick up a large exfoliated hBN flake by a PDMS droplet (Fig.\,\ref{fig:sample}a i.). The hBN flake should be large enough to cover the whole active area of the spin-valve to reduce the risk of solvent penetration and to prevent airborne contaminants and oxidation of the electrodes (see optical image in Fig.\,\ref{fig:sample}e).\cite{drogeler_12ns_2016} Then, the pre-etched hBN/CVD-BLG bar is picked up with this large hBN flake (Fig.\,\ref{fig:sample}a ii.). During this process, the stamp is first put into contact with the stack at a temperature of \SI{135}{\celsius}. The PDMS droplet allows to apply a local force between the hBN transfer flake and the hBN/CVD-BLG bar, enhancing their adhesion and allowing a reliable transfer. Before the stamp can be lifted from the substrate with the stack attached to it, the polymer has to be cooled down to $T\leq \SI{75}{\celsius}$ to increase the lift-up probability.

In a next step, the entire stack is placed onto pre-defined Co(\SI{30}{\nano\meter})/\ch{Al2O3}(\SI{1}{\nano\meter}) electrodes fabricated by e-beam evaporation (Fig.\,\ref{fig:sample}a iii.).\cite{Volmer2021Mar} During the transfer process, we increase the mechanical pressure of the PDMS droplet tool to improve the contact to the \ch{Al2O3}/Co electrodes. It is expected that the high local forces achieved by the PDMS stamping tool improve the electrical contact to the \ch{Al2O3} tunnel barrier. Furthermore, the transparency of the PDMS droplet tool and the small contact area allow to precisely align and deposit the final hBN/hBN/CVD-graphene stack onto the electrodes with a spatial precision of $\leq \SI{1}{\micro\meter}$. 

The stack with the PC foil is finally released at a temperature of \SI{180}{\celsius} while continually retracting the PDMS droplet (Figs.\,\ref{fig:sample}a iv.\,and \ref{fig:sample}e). We point out that the PDMS droplet tool can easily be cleaned and reused many times. In particular, it is not necessary to dissolve the PC membrane fused to the substrate. Accordingly, our transfer method allows the precise stacking and transfer onto small areas resulting in ready-to-use devices without potential contaminations by a wet-chemical removal of the polymer (see Ref.\,\citenum{drogeler_12ns_2016} where we demonstrate that solvents can partially lift heterostructures and potentially contaminate the graphene that was meant to be protected by the covering hBN layer). 

All transport measurements were performed at room temperature under vacuum conditions using a standard low-frequency lock-in technique.\cite{Volmer_contact-induced_2015} A typical gate dependent graphene resistance curve is shown in Fig.\,\ref{fig:measurements}a for the suspended BLG region A in Fig.\,\ref{fig:sample}e. The gate voltage was swept from negative to positive voltages (trace, dotted line) and back (retrace, solid line) in a range of \SIrange{-40}{40}{\volt}. We find the charge neutrality point (CNP) at \SI{-4}{\volt} for the trace and at \SI{-2}{\volt} for the retrace. This very small hysteresis and the low residual doping at $V_G=0\si{\volt}$ demonstrate that our fabrication technique yields highly clean devices. 

\begin{figure}[tb]
\includegraphics[width=0.45\textwidth]{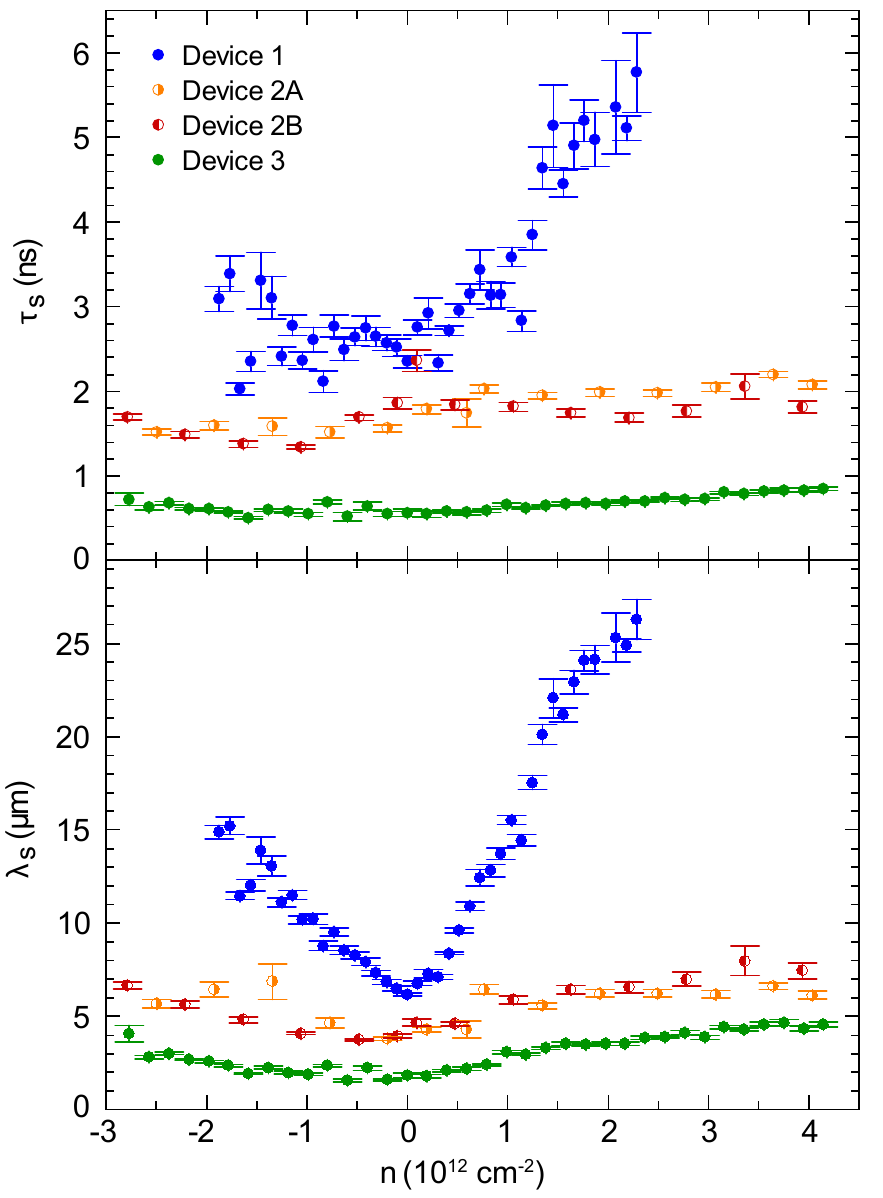}
\caption{Tunability of the room temperature spin lifetimes (upper panel) and diffusion lengths (lower panel) with the charge carrier density $n$ for three different CVD bilayer graphene spin-valve devices fabricated with the droplet tool.}
\label{fig:taulambda}
\end{figure}

The charge carrier mobility $\mu$ is extracted from the gate dependent conductance $\sigma$ with $\mu=1/e\cdot (\Delta\sigma/\Delta n)$ yielding $\mu=\SI{23800}{\square\cm(\volt\second)^{-1}}$ at an electron density of $n=\SI{1E12}{\per\centi\metre\squared}$.

This value is more than a factor of two larger than previous values obtained on inverted BLG spin-valves fabricated from exfoliated bilayer graphene\cite{drogeler_nanosecond_2014} and equals the best room temperature values of single-layer graphene inverted spin-valves.\cite{drogeler_12ns_2016} In BLG, higher mobilities were achieved only in fully-encapsulated devices with 1D side contacts for charge transport experiments.\cite{Schmitz2017Jun}

Next, we focus on the non-local spin transport measurements, where we used the non-local 4-terminal configuration. Contacts C1 and C2 act as source and drain electrodes while contacts C3 and C4 are used to probe the non-local spin voltage (see Fig.\,\ref{fig:sample}e). Hanle spin precession measurements were performed to determine the spin lifetime $\tau_s$ and the spin diffusion length $\lambda_s=\sqrt{D_s \tau_s}$, where $D_s$ is the spin diffusion coefficient. Fig.\,\ref{fig:measurements}b shows Hanle spin precession measurements obtained from region A of the device in Fig.\,\ref{fig:sample}e at an electron density of \SI{2E12}{\per\centi\metre\squared} measured for both parallel (blue curve) and antiparallel (red curve) alignments of the respective spin injection electrode C2 and spin detection electrode C3. Fitting these curves yields a spin lifetime of $\tau_s=\SI{5.1}{\nano\second}$ and a remarkably long spin diffusion length of \SI{24.9}{\micro\metre}. 

We note that the fitting was done by using a stationary solution of the Bloch-Torrey equation obtained from Ref.\,\citenum{johnson_coupling_1988} (see Ref.\,\citenum{fabian_semiconductor_2007} for a more detailed derivation) with an additional parabolic background term.\cite{Volmer_contact-induced_2015} However, the fact that the resulting spin diffusion lengths are comparable to the device dimensions (compare to Fig.\,\ref{fig:sample}e) seemingly contradicts a fundamental assumption in the aforementioned derivation of the fit function: The outer electrodes are assumed to be placed in a region of the device in which the induced spin accumulation of the inner injection electrode is reduced to a negligible level by relaxation processes. Recent studies have predicted that otherwise our used fitting function yields erroneous results for both the fitted spin lifetime and spin diffusion length.\cite{Vila_Roche_2020,Fourneau2021,Droegeler-PSSB-2017}

However, as we show and discuss in the Supporting Information, the spin accumulation does not reach the outermost contacts even though the large spin diffusion length would suggest otherwise. 
We attribute this to contact-induced spin relaxation processes caused by the contacts that lie between the measured graphene region and the outermost contacts.\cite{Maassen2012,Volmer_Suppression_2014,Idzuchi-Revisiting-2015,Amamou-ContactInduced-2016,Stecklein-ContactInduced-2016,Droegeler-PSSB-2017} For the case that contact-induced spin relaxation plays a significant role, many studies predict that the fitting equation we are using (obtained from Ref.\,\citenum{johnson_coupling_1988}) actually underestimates the spin lifetime as well as the spin diffusion length \cite{Droegeler-PSSB-2017,Maassen2012,Amamou-ContactInduced-2016,Idzuchi-Revisiting-2015,Stecklein-ContactInduced-2016}. Therefore, the values presented in this study are likely conservative lower bounds to the real spin transport parameters. Nevertheless, we chose to continue using this conservative approach to ensure that the results of this study can be easily compared to both our previous studies and many other publications where the same fitting function can be found (see a more detailed discussion about our fit function in the Supporting Information).

Next, we take a look at the spin amplitude $\Delta R_{\mathrm{nl}}=R_{\mathrm{nl}}^{\mathrm{p}}-R_{\mathrm{nl}}^{\mathrm{ap}}$, which can be extracted from the Hanle curves at $B=\SI{0}{\tesla}$ for both parallel (p) and antiparallel (ap) magnetization configurations of the electrodes (see Fig.\,\ref{fig:measurements}c). 
We note that the non-local resistances $R_{\mathrm{nl}}^{\mathrm{p,ap}}$ are given by the ratio $R_{\mathrm{nl}}=V_{\mathrm{nl}}/I_{\mathrm{ac}}$ of the non-local voltages $V_{\mathrm{nl}}$ between contacts C3 and C4 and the applied local current between contacts C1 and C2, which typically is $I_{\mathrm{ac}}=\SI{5}{\micro\ampere}$. 
The spin amplitude remains large for both electron ($n>0$) and hole ($n<0$) doping indicating that the droplet-assisted dry transfer method can yield high spin injection and detection efficiencies. 
The corresponding contact-resistance-area products are in the range of \SIrange{35}{430}{\kilo\ohm (\micro\metre)\square}. 
Such high contact resistances can decrease contact-induced spin scattering to such extents that nanosecond spin lifetimes are achievable.\cite{Volmer_Suppression_2014,Droegeler-PSSB-2017,Maassen2012,Amamou-ContactInduced-2016,Idzuchi-Revisiting-2015,Stecklein-ContactInduced-2016} Nevertheless, as we discuss in the Supporting Information our contacts are not ideal and contact-induced spin relaxation is likely still the bottleneck for the overall spin transport performance. 

The dependence of the spin lifetimes and the spin diffusion lengths on the charge carrier density at room temperature are shown in Fig.\,\ref{fig:taulambda}. Next to the blue data points of device 1 that was discussed in Figs.\,\ref{fig:sample} and \ref{fig:measurements} we also include measurements from two additional devices, which were fabricated by the same PDMS droplet method (Raman measurements, electrical transport data, and raw spin transport measurements on these additional devices are shown in the Supporting Information). While all devices show long room temperature spin lifetimes in the nanosecond range, only device A exhibits the strong V-shape dependence of the spin transport parameters, reaching a spin lifetime of \SI{5.8}{\nano\second} and a spin diffusion length of \SI{26.3}{\micro\metre} at the largest electron density of around \SI{2.3E12}{\per\centi\metre\squared}. The gate dependent increase in the spin transport parameters at these carrier densities is similar to the best iSVs made from exfoliated SLG, where spin lifetimes of \SI{12.6}{\nano\second} and spin diffusion lengths of \SI{30.5}{\micro\metre} were achieved at much larger carrier densities of \SI{4.7E12}{\per\centi\metre\squared}. 
Device~1 outperforms all reports on spin transport in BLG spin-valves and shows a doubling of both spin lifetime and spin diffusion length compared to the best previous devices.\cite{Ingla_88percent_2016} 

Device~2 also shows excellent spin transport properties at the charge neutrality point ($n=0$) reaching \SI{2}{\nano\second}, which is only slightly below the respective value for device~1. 
In this device, there are two neighboring spin transport regions showing identical values for the spin lifetime as well as for the spin diffusion length (red and orange data points in Fig.\,\ref{fig:taulambda}), indicating the lateral homogeneity of the transport properties. 
While both the contact resistances and the non-local spin resistance are similar to device A, there is nearly no gate tunability in the spin transport parameters. 
Although the absence of gate tunability is currently not understood, we note that we also observed this behavior in some of our previous graphene-based iSVs.\cite{drogeler_nanosecond_2014,drogeler_nanosecond_2015,drogeler_12ns_2016} 
Rather weak gate tunability is also seen in device~3 (green data points in Fig.\,\ref{fig:taulambda}). While the transfer of the hBN/CVD-BLG bar has worked equally well as for the other devices, the overall spin transport performance is found to be lower. 
We attribute this to small contact resistances of below \SI{20}{\kilo\ohm(\micro\metre)\square} and an overall small non-local spin resistance of \SI{0.4}{\ohm} (at similar electrode spacings compared to the other devices), indicating that contact-induced spin scattering might be the reason for the reduced device performance (see a more detailed discussion in the Supporting Information).

The integration of CVD-grown BLG into spintronic devices promises three distinct advantages over exfoliated SLG: First, the possibility to scale up device dimensions up to wafer-scale.
Second, the incorporation of a tunable bandgap to include semiconductor features. 
Third, a predicted larger proximity-induced tuning of the spin transport properties. 
So far, however, BLG did not meet these expectations as both the reported values of spin lifetimes and charge carrier mobilities fall significantly short when compared to SLG. 

We have demonstrated that this is related to the process technologies used to date.
In this work, we instead use a novel droplet-assisted transfer method that compensates for those previous process drawbacks in device fabrication. 
This leads to lateral CVD-BLG spin valves that are competitive to state-of-the-art exfoliated SLG devices.

In gate-dependent Hanle measurements, we regularly achieve spin lifetimes of \SIrange{1}{2}{\nano\second} and, for the best sample, up to \SI{5.8}{\nano\second} at room temperature. The respective spin diffusion lengths go up to \SI{26}{\micro\metre} and the charge carrier mobility reaches up to \SI{24000}{\square\cm(\volt\second)^{-1}}. 
In terms of spin transport parameters, this corresponds to at least a doubling of previously published BLG samples including those from exfoliated BLG.
Therefore, we expect that our result will pave the way to high-quality scalable devices with integrated tunable proximity and semiconductor properties.

\begin{acknowledgements}
The authors thank S.\,Staacks for help on the figures. This project has received funding from the European Union's Horizon 2020 research and innovation programme under grant agreement No.\,881603 (Graphene Flagship) and the Deutsche Forschungsgemeinschaft (DFG, German Research Foundation) through DFG (BE 2441/9-1), and by the Helmholtz Nano Facility (HNF)\cite{HNF} at the Forschungszentrum J\"ulich. Growth of hexagonal boron nitride crystals was supported by the Elemental Strategy Initiative conducted by the MEXT, Japan, Grant Number JPMXP0112101001, JSPS KAKENHI Grant Numbers JP20H00354, and the CREST(JPMJCR15F3), JST.
\end{acknowledgements}

\bibliographystyle{apsrev4-2}
\bibliography{ms.bbl}

\end{document}


\author{Timo Bisswanger}
\author{Zachary Winter}
\author{Anne Schmidt}
\author{Frank Volmer}
\affiliation{2nd Institute of Physics and JARA-FIT, RWTH Aachen University, 52074 Aachen, Germany}
\author{Kenji Watanabe}
\affiliation{Research Center for Functional Materials,
National Institute for Materials Science, 1-1 Namiki, Tsukuba 305-0044, Japan}
\author{Takashi Taniguchi}
\affiliation{International Center for Materials Nanoarchitectonics,
National Institute for Materials Science, 1-1 Namiki, Tsukuba 305-0044, Japan}
\author{Christoph Stampfer}
\affiliation{2nd Institute of Physics and JARA-FIT, RWTH Aachen University, 52074 Aachen, Germany}
\affiliation{Peter Gr\"unberg Institute (PGI-9) Forschungszentrum J\"ulich, 52425 J\"ulich, Germany}
\author{Bernd Beschoten}
\email{bernd.beschoten@physik.rwth-aachen.de}
\affiliation{2nd Institute of Physics and JARA-FIT, RWTH Aachen University, 52074 Aachen, Germany}

\title{Supporting Information: CVD bilayer graphene spin valves with 26 \si{\micro\metre} spin diffusion length at room temperature}
\maketitle

\tableofcontents

\begin{figure*}[tb]
\includegraphics{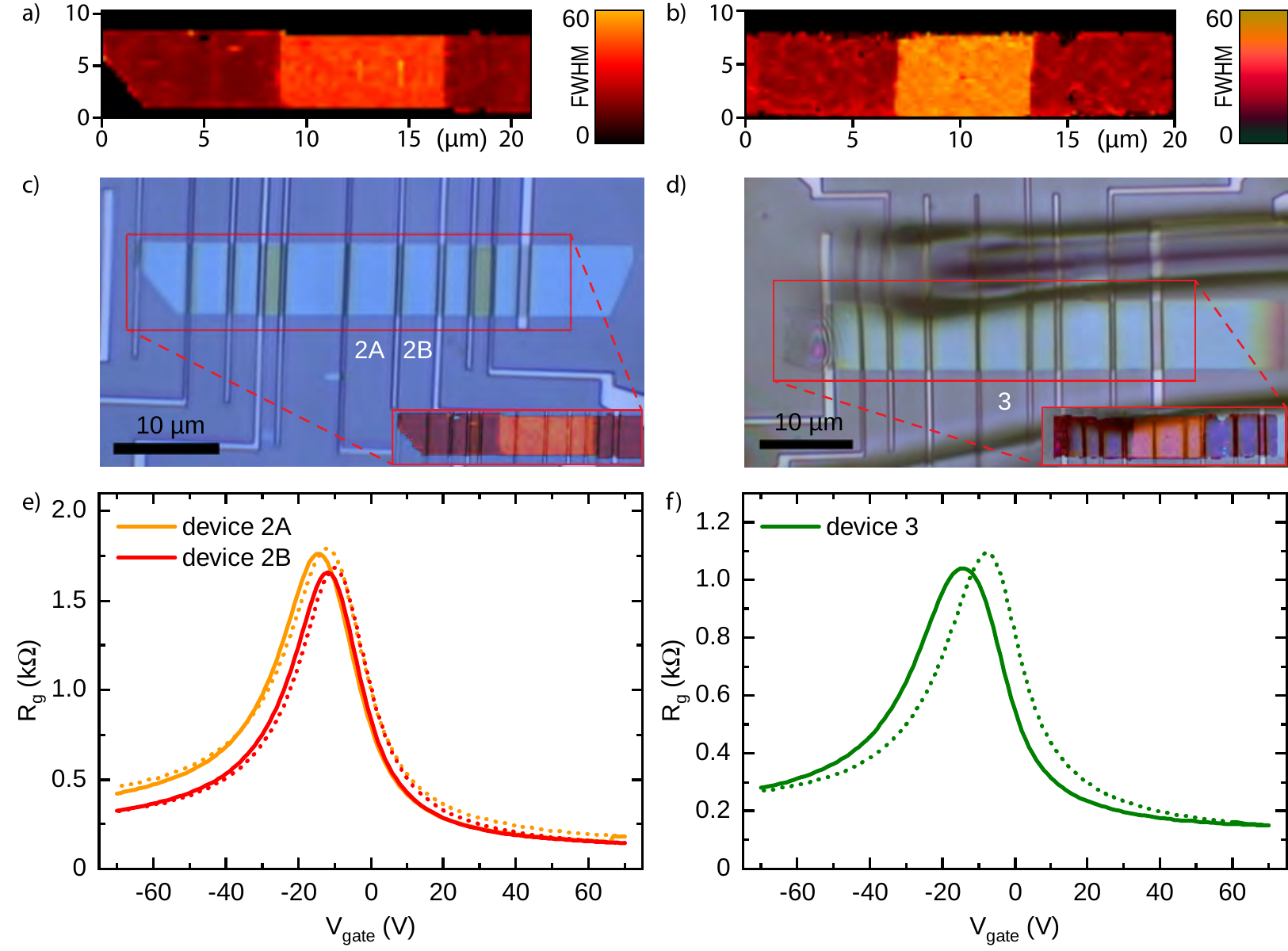}
\caption{a),b) Map of the Raman \textit{2D}-mode linewidth from device 2 (a) and device 3 (b).  c), d) Optical images of device 2 and 3, respectively. The insets show an overlay between the optical images and the Raman maps, clearly showing the positions of the BLG regions in the devices. The optical images were taken while the devices are still covered with a PC polymer film, resulting in some optical distortions. e),f) Gate dependent graphene resistances of the investigated regions measured at room temperature.}
\label{fig:S1}
\end{figure*}

\section{Raman measurements and electrical transport data of other devices}
In this section, we present additional information on the sample structure, confocal Raman spectroscopy data and charge transport measurements of device 2 (Figs.\,\ref{fig:S1}a,c,e) and device 3 (Figs.\,\ref{fig:S1}b,d,f).

The Raman maps in Figs.\,\ref{fig:S1}a and b show the spatially resolved Raman \textit{2D}-linewidth, which is the most significant indicator to distinguish bilayer graphen (BLG) from single layer graphene (SLG).
The BLG areas can be well identified as areas of increased linewidths. With the help of an overlay between optical image and Raman map (see insets in Figs.\,\ref{fig:S1}c,d) we can clearly identify the regions in each device that consist of BLG. The optical images were taken while the devices are still covered with a PC polymer film, resulting in some optical distortions.

The gate dependent graphene resistances at room temperature are depicted in Figs.\,\ref{fig:S1}e and f. Shown are both sweep directions of the gate voltage (solid vs dashed lines).
The resulting carrier mobilities at room temperature in device 2 are $\mu_\mathrm{2A}=\SI{14400}{\square\cm(\volt\second)^{-1}}$ and $\mu_\mathrm{2B}=\SI{14700}{\square\cm(\volt\second)^{-1}}$ at a charge carrier density of $n=1.6\times 10^{12}$, and the mobility in device 3 is $\mu_\mathrm{3}=\SI{10800}{\square\cm(\volt\second)^{-1}}$ at $n=1.4\times 10^{12}$.
The nearly identical mobilities in the two regions of device 2 indicate the homogeneity of the sample in the BLG region, which is also seen in the spin transport measurements (see main manuscript).

\section{Discussion of the fit function and the reliability of the fitting procedure}
\begin{figure*}[tb]
\includegraphics{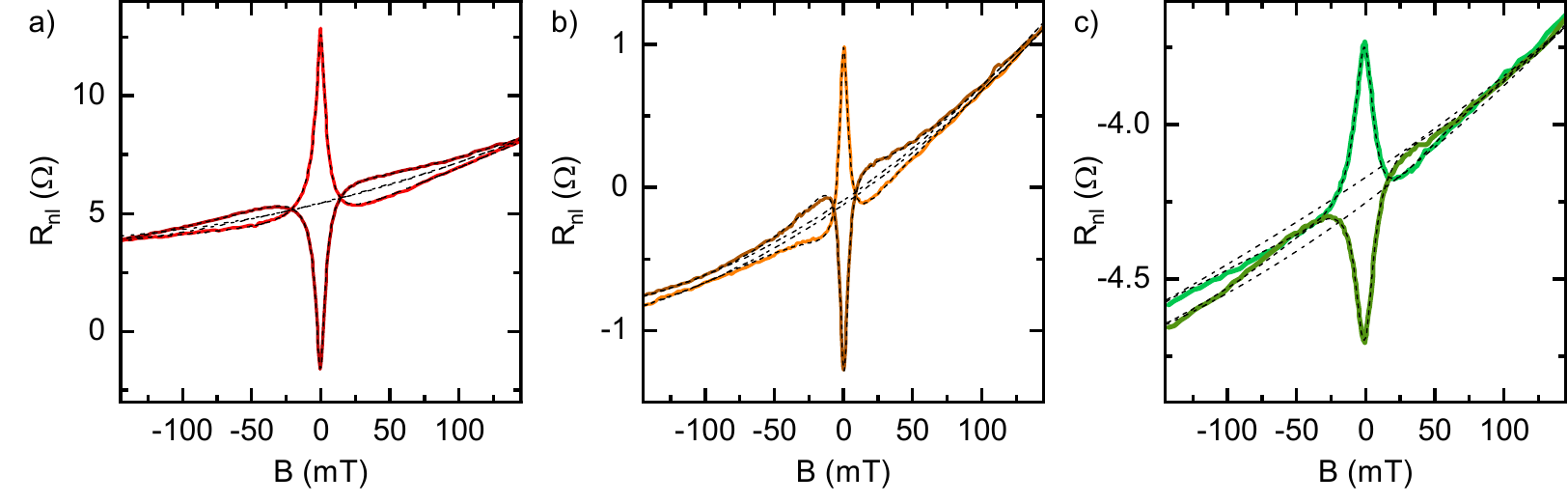} 
\caption{Hanle spin precession measurements in parallel (bright) and anti-parallel (dark) contact configuration of devices 2A,  2B, and 3 from left to right. The black dashed lines represent the parabolic background taken into account by the fit routine as well as the total fit curve according equation\,\ref{backgroundfit}. The curves were taken at zero backgate voltage for device 2 and at 70\,V for device 3.}
\label{fig:SI_Hanle}
\end{figure*}

To extract spin lifetimes and spin diffusion coefficients from Hanle spin precession measurements, we first consider the stationary Bloch-Torrey equation:
\begin{equation}
	\frac{\partial \vec{s}}{\partial t}\;=\;\vec{s}\times \vec{\omega}_0+D_{\mathrm{s}}\nabla^2\vec{s}-\frac{\vec{s}}{\tau_{\mathrm{s}}}\;=0,
	\label{Bloch-Torrey-equation}
\end{equation}
where $\vec{s}$ is the net spin vector, $\vec{\omega_0}=g\mu_{B} \vec{B}/\hbar$ is the Larmor frequency (we assume $g=2$), $D_{\text{s}}$ is the spin diffusion constant, and $\tau_{\text{s}}$ is the spin lifetime. With $L$ being the distance between spin injection and spin detection electrodes, we define the following dimensionless parameters:
\begin{equation}
    b=\omega_0\tau_{\mathrm{s}}=-\frac{g\mu_B B_\perp}{\hbar}\tau_{\mathrm{s}},
\end{equation}
\begin{equation}
    f(b)=\sqrt{1+\sqrt{1+b^2}},
\end{equation}
\begin{equation}
    l=\sqrt{\frac{L^2}{2\tau_{\mathrm{s}}D_{\mathrm{s}}}}.
\end{equation}
With the help of these parameters, equation~\ref{Bloch-Torrey-equation} can be solved analytically under the following main assumptions: 1.)\:The equation is considered only in one dimension, 2.)\:there is neither a spatial variation nor an anisotropy in $\tau_{\text{s}}$ and $D_{\text{s}}$, 3.)\:injection and detection of spins occur at point-like, ideal contacts, and 4.)\:the outer electrodes of the non-local measurement geometry are placed in a region of the device in which the induced spin accumulation of the inner injection electrode is reduced to a negligible level by relaxation processes. Under these assumptions, the analytical solution to equation~\ref{Bloch-Torrey-equation} is:\cite{johnson_coupling_1988,fabian_semiconductor_2007}
\begin{equation}
  R_{\textrm{nl}}^{\textrm{Hanle}}(B) = \pm \frac{\Delta R_{\textrm{nl}}}{2} \cdot F(b,l),
\label{Hanle}
\end{equation}
with:
\begin{equation}\label{EqFourneau}
\begin{split}
    F(b,l)=\frac{1}{f^2(b)-1}
    \cdot\left[ f(b)\cos\left(\frac{lb}{f(b)}\right)\right.\\\left.-\frac{b}{f(b)}\sin\left(\frac{lb}{f(b)}\right)\right]
    \cdot \exp(-lf(b)).
    \end{split}
\end{equation}
The sign in equation~\ref{Hanle} depends on either a parallel or an antiparallel alignment between the magnetization of injection and detection electrodes. However, in addition to the theoretically expected Hanle spin precession signal, there is an additional background signal present in real measurements (see e.g.\,Fig.\,\ref{fig:SI_Hanle}). This background can be well fitted by a polynomial function of second order with coefficients $c_i$ and can be explained by two effects: 1.)\:A Hall-effect related signal caused by a spatially inhomogenous injection and detection of spins and 2.)\:measurement artifacts caused by a non-zero common-mode voltage.\cite{Volmer_contact-induced_2015,Volmer-arxiv-2021} The Hanle spin precession measurements are thus fitted by the following equation:
\begin{equation}
\label{backgroundfit}
    R_{\textrm{nl}}^{\textrm{total}}(B) \;=\; \pm \frac{\Delta R_{\textrm{nl}}}{2} \cdot F(b,l) + c_2 B^2 + c_1 B + c_0.
\end{equation}
An asymmetric portion in the Hanle curve due to imperfectly aligned samples or slightly tilted domains can be accounted for by an appropriate term of the solution of equation~\ref{Bloch-Torrey-equation} for crossed injection and detection.\cite{drogeler_12ns_2016}

In the last few years quite some studies pointed out that the assumptions, that are necessary to derive the analytical expression in equation~\ref{Hanle}, may not be applicable to real devices. A wide variety of different fit functions were proposed, each incorporating different aspects like: 1.)\:Contact-induced spin relaxation, 2.)\:regions with different spin transport properties, 3.)\:spin lifetime anisotropy, 4.)\:the effect of finite device dimensions, 5.)\:the impact of localized states, or 6.)\:the transition from a diffusive to a ballistic spin transport regime that may be encountered in devices with high charge carrier mobilities.\cite{Maassen2012,Idzuchi-Revisiting-2015,Amamou-ContactInduced-2016,Stecklein-ContactInduced-2016,Vila_Roche_2020,Fourneau2021,Guimaraes_Suspended_2012,Maassen-LocalizedStatesInfluence-2013,vandenBerg-ObservationAnomalousHanle-2015,Ingla_88percent_2016,Droegeler-PSSB-2017,Zhu_modeling_2018}

This raises the question which kind of fit function should be used to extract the genuine spin lifetimes of graphene from Hanle spin precession measurements. In this respect, it is important to note that the motivation behind many of the abovementioned alternative fit functions is the fact that equation~\ref{Hanle} actually underestimates the extracted spin lifetimes in many cases (the only exceptions will be discussed further below). Therefore, using equation~\ref{Hanle} gives in most cases a conservative lower bound, making any statements of long spin lifetimes or spin diffusion constants more robust.

Another important reason why we stick to equation~\ref{Hanle} is that many of the proposed fit functions have at least one of the two following problems: 1.)\:Considering additional effects inevitably necessitates a more complex mathematical expression than equations~\ref{Hanle} and \ref{EqFourneau}, which makes it difficult to fit the Hanle spin precession curves in a reproducible and reliable way.
Some of the abovementioned studies even lack an analytical solution at all, so that the experimental data must be analyzed by simulating the Hanle spin precession curves by varying the assumed spin transport parameters (cf.\,e.g.\,Refs.\,\citenum{Guimaraes_Suspended_2012, Ingla_88percent_2016}). 2.)\:Many of the alternative fit functions are derived by assumptions that are unlikely met in real devices.
Some studies e.g.\,assume homogeneous transport properties over the whole device, although it was shown that contacts highly impact the properties of graphene.\cite{drogeler_nanosecond_2014,Amamou-ContactInduced-2016,LeeEduardo2008,Gong2013a,McCreary2011} Other studies describe contact-induced spin relaxation processes via a single parameter, i.e.\,a contact resistance. This approach necessitates a homogeneous contact-to-graphene interface, although it was shown that the contact characteristics are often dominated by pinholes.\cite{drogeler_12ns_2016,Volmer_contact-induced_2015,Volmer_Suppression_2014}

Overall, none of the abovementioned alternative fit function is using a model that describes a realistic device in each detail (i.e.\,contact-induced spin relaxation, regions with different spin transport properties, spin lifetime anisotropy, the effect of finite device dimensions, and so on).
Instead of cherry-picking a fit function that is incorporating only one additional effect (especially without knowing if this additional effect is even the one that is affecting the spin transport properties in the device the most) we think it is still the best approach to use the long-established fit function of equations~\ref{Hanle} and \ref{EqFourneau}. Especially as this fit function normally yields conservative lower bounds for the extracted spin lifetimes.

\begin{figure}[tb]
\includegraphics[width=\linewidth]{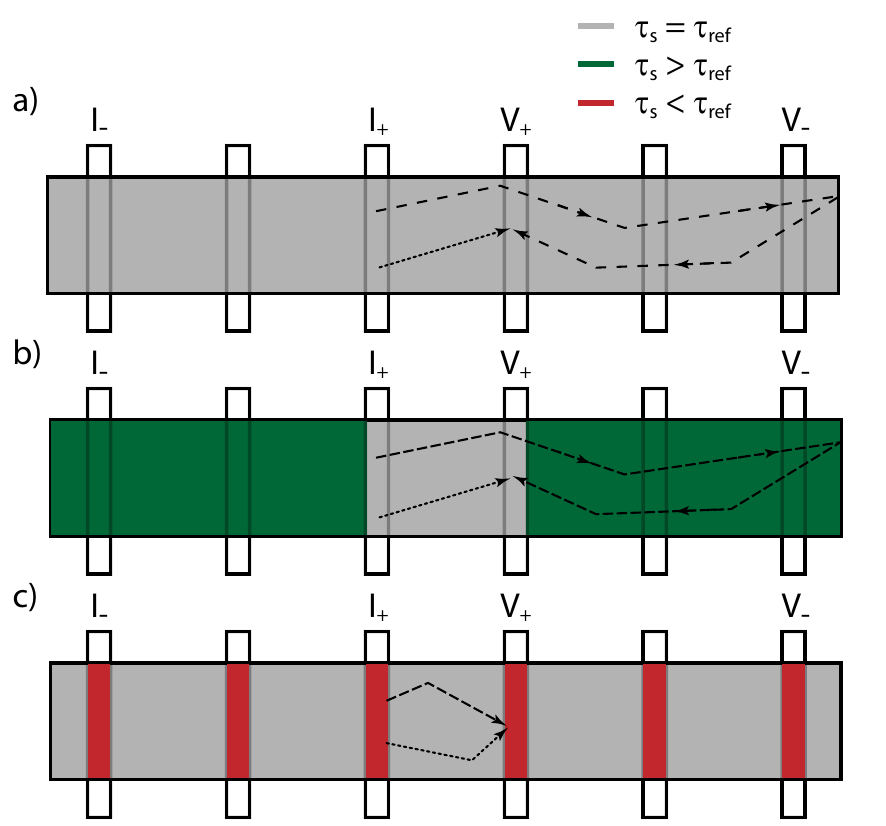}
\caption{Schematic representations of devices with spatially varying spin lifetimes. a) The spin lifetime is spatially homogeneous along the whole device. b) The spin lifetime is higher in the outer parts of the channel (SLG in our devices) compared to the inner part (BLG in our devices). c) The spin lifetimes is highly diminished in the parts of graphene that is in contact to the electrodes.}
\label{fig:SI_Fit}
\end{figure}

In the following we discuss special cases in which equation~\ref{Hanle} may in fact overestimate the spin lifetimes. For this we refer to Fig.\,\ref{fig:SI_Fit} that is showing schematic representations of graphene flakes (horizontal rectangles) that are transferred on top of pre-fabricated electrodes (smaller, vertical rectangles). In this figure we color-coded regions which have a lifetime equal to a reference value (grey), regions with much longer spin lifetimes than the reference value (green), and regions with much shorter lifetimes (red).

In one of our previous publications we have demonstrated by finite-element simulations that an overestimation of the fitted spin lifetime can occur if the following two conditions are met:\cite{Droegeler-PSSB-2017} 1.)\:The spin transport properties are homogeneous over the whole device area as seen in Fig.\,\ref{fig:SI_Fit}a and 2.)\:the spin diffusion length is comparable to the length of the device. Under such conditions, a spin polarized charge carrier can reach the end of the graphene flake where it will be reflected and therefore can once again reach the detection electrode (dashed line in Fig.\,\ref{fig:SI_Fit}a). The length of the corresponding diffusion path is much longer compared to a spin that is absorbed by the detection electrode right away (pointed line in Fig.\,\ref{fig:SI_Fit}a). When a magnetic field is applied perpendicular to the graphene sheet, the spin on the first path will acquire a larger phase compared to the spin on the second path. As the recorded Hanle spin precession curve is the superposition of a whole ensemble of spins that has undergone a variety of different paths, the overall spin accumulation underneath the detector will therefore dephase faster with increasing magnetic field if reflection at the end of the graphene flake becomes relevant. This leads to a narrower Hanle curve which mimics longer spin lifetimes.

The resulting overestimation of the spin lifetime is even more pronounced, if the neighboring regions exhibit longer spin lifetimes (see green regions in Fig.\,\ref{fig:SI_Fit}b). This is due to the fact that under this condition it is even more likely for spin-polarized charge carriers to be reflected at the end of the graphene sheet and coming back to the electrode before their spin polarization is lost because of scattering events. As the outermost regions in our devices consist of SLG, whereas the measured region in the center consists of BLG, the question arises if the condition of Fig.\,\ref{fig:SI_Fit}b is actually met. However, we note that in all of our previous studies, there is no clear indication that the spin transport properties are more favorable in SLG compared to BLG.\cite{drogeler_nanosecond_2014,drogeler_nanosecond_2015,drogeler_12ns_2016,drogeler_dry-transferred_2017}

More importantly, we demonstrated in Ref.\,\citenum{Droegeler-PSSB-2017} that the overestimation of the spin lifetime under conditions like the ones in Figs.\,\ref{fig:SI_Fit}a,b is accompanied by a significant underestimation of the spin diffusion coefficient. As a result, we observed in our simulations that the fitted spin diffusion length decreases away from the charge neutrality point, although we put an increasing spin diffusion length into the simulation (see Fig.\,1f in Ref.\,\citenum{Droegeler-PSSB-2017} where the solid line represents the values put into the simulation and the data points are the fitted values). This is in strong contradiction to the gate dependence of the fitted spin diffusion lengths in our devices, which show a clear increase of the spin diffusion length towards higher charge carrier densities (see Fig.\,3 in the main manuscript).

\begin{figure*}[tb]
\includegraphics{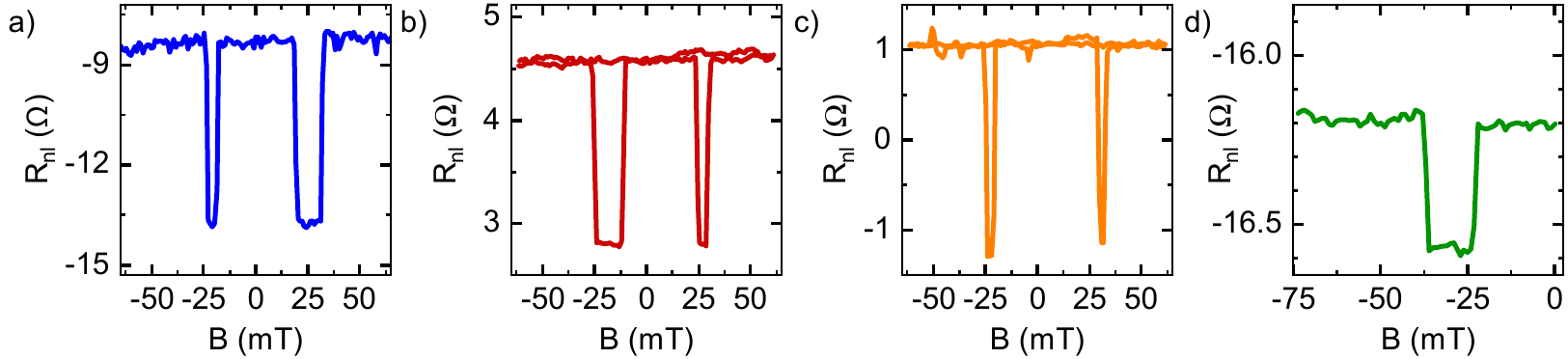}
\caption{Spin valve measurements of devices 1, 2A, 2B, and 3 from left to right. All measurements were taken at zero backgate voltage.}
\label{fig:SI_SV}
\end{figure*}

Finally, it is very important to note that the situations in Figs.\,\ref{fig:SI_Fit}a,b would result in the fact that the spin accumulation that is injected by $I_+$ would reach the outer detection electrode $V_-$. Additionally, the spin accumulation created by $I_-$ would also reach $V_+$. Therefore, the switching of the outer electrodes should show up in the spin valve measurements as multiple switching events.\cite{Popinciuc2009,Berger2015,Fourneau2021} However, we do not observe multiple switching in any of our devices indicating that only the magnetization reversal of the  innermost electrodes ($V_+$ and $I_+$) can be observed in our spin valve measurements (see Fig.\,\ref{fig:SI_SV}).

Fully consistent to this observation is the fact that we also do not observe any impact of the outermost contacts to the Hanle spin precession measurements. For this we adapt the fit function that is proposed in Ref.\,\citenum{Fourneau2021}, which the authors suggest to be used if the spin diffusion length is in the same range as the device's dimension:
\begin{equation}\label{EqFourneauMain}
    \begin{split}
    R_{\mathrm{nl}}=R_0
    [F(L_{ch},B_\perp)
    -F(L_1+L_{ch},B_\perp)\\
    -F(L_2+L_{ch},B_\perp)+F(L_1+L_2+L_{ch},B_\perp)
    ],
    \end{split}
\end{equation}
with $L_{1}$ being the distance between $I_+$ and $I_-$, $L_{2}$ being the distance between $V_+$ and $V_-$, and $L_{ch}$ being the distance between $I_+$ and $V_+$. See Ref.\,\citenum{Fourneau2021} for a more detailed discussion of the factor $R_0$. However, the study in Ref.\,\citenum{Fourneau2021} assumes identical spin injection and detection efficiencies for each contact, which is not realistic in real devices. We therefore slightly modified the proposed fit function by assigning separate spin injection and detection efficiencies $P_i$ to each contact:
\begin{equation}\label{EqFourneauMod}
    \begin{split}
    R_{\mathrm{nl}}=R_0[P_{I+} P_{V+} F(L_{ch},B_\perp)-P_{I-} P_{V+} F(L_1+L_{ch},B_\perp) \\
    - P_{I+} P_{V-} F(L_2+L_{ch},B_\perp) + P_{I-} P_{V-} F(L_1+L_2+L_{ch},B_\perp)].
    \end{split}
\end{equation}
If we use equation~\ref{EqFourneauMod} to fit of our Hanle spin precession curves, the fit routine always sets the spin injection and detection efficiencies of the outermost contacts to zero (i.e.\,$P_{I-} = P_{V-} =0$). This means that the fitting routine cannot detect any contributions of the outermost contacts to the Hanle spin precession measurements.

These observations (i.e.\,no signal from the outermost electrodes and no severe underestimation of the spin diffusion coefficient) show that our devices do not meet the conditions that are illustrated in Figs.\,\ref{fig:SI_Fit}a,b. Instead, the spin polarization must relax before it can reach the ends of the graphene channel. The most likely explanation for this is that the areas of the graphene flake that are in contact with the electrodes exhibit a strongly reduced spin lifetime (see Fig.\,\ref{fig:SI_Fit}c). As we discuss in the next section, we believe that contact-induced spin relaxation is in fact the main bottleneck for the spin performance of our devices. With respect to the fitting, it is important to note that we have already demonstrated that contact-induced spin relaxation underestimates the spin lifetimes obtained from equation~\ref{Hanle}.\cite{Droegeler-PSSB-2017}

\section{Contact-induced spin relaxation as the bottleneck of device performance}
As discussed in the last section, the spin accumulation does not reach the outermost contacts in our devices, although the spin diffusion length in the region between electrodes $I_+$ and $V_+$ (see Fig.\,\ref{fig:SI_Fit}c) is in the range of the overall length of the device. We believe that this observation can be explained by contact-induced spin relaxation processes.\cite{Maassen2012,Amamou-ContactInduced-2016,Droegeler-PSSB-2017,Idzuchi-Revisiting-2015,Stecklein-ContactInduced-2016,Volmer_Suppression_2014} It is known that contacts can significantly modify the electrical characteristics of graphene.\cite{Giovannetti-DopingGrapheneMetal-2008,Gong2013a,LeeEduardo2008} One reason for this is a hybridization between graphene with many metals.\cite{Varykhalov2012,Varykhalov-EffectNoblemetalContacts-2010,Voloshina-GrapheneMetallicSurfaces-2012,Hsu-SurfaceInducedHybridizationGraphene-2014} Although a dielectric tunnel barrier between the graphene and the metal can in principal prevent such a hybridization effect, there is always the problem of pinholes in such barriers,\cite{Volmer_Suppression_2014,Volmer_contact-induced_2015,drogeler_12ns_2016} at which positions metal and graphene can still interact. Of course, there are studies claiming a pinhole-free fabrication of tunnel barriers on graphene, e.g.\,the one in Ref.\,\citenum{Dlubak-HomogeneousPinholeFree-2012}. However, the same group also reported that with the same fabrication method defects are introduced into graphene.\cite{Dlubak-AreAl2O3MgO-2010} Then, there is a study claiming atomically smooth tunnel barriers by using a seed layer of titanium.\cite{Wang-GrowthAtomicallySmooth-2008} However, the same group demonstrated that even a sub-monolayer of titanium significantly modifies the electronic properties of graphene and that the sub-monolayer of titanium takes hours to completely oxidize,\cite{McCreary2011} which is most likely due to a hybridization between titanium and graphene.\cite{Hsu-SurfaceInducedHybridizationGraphene-2014}

Overall, it is the highly inert nature of the chemical bonds in graphene that seems to result in only two possible outcomes when fabricating contacts on top of graphene: Either a fabrication method is used that is introducing artificial nucleation sites or modifying the graphene layer to overcome the inert nature of graphene,\cite{Wang-GrowthAtomicallySmooth-2008,Dlubak-AreAl2O3MgO-2010} which most likely impacts the electrical and spin transport properties of graphene; or the dielectric layer is deposited on the pristine graphene flake, which leads to island growth conditions and therefore tunnel barriers with very rough interfaces and possible pinholes, both possible causes of spin relaxation processes.\cite{Txoperena-SpinInjectionLocal-2016,Dash-SpinPrecessionInverted-2011,Muduli-LargeLocalHall-2013,Volmer_Suppression_2014}

In fact, our fabrication method, in which we transfer a graphene flake on top of pre-fabricated electrodes, was designed to overcome the problem of the challenging growth of tunnel barriers on top of graphene: The growth of first cobalt and then an insulating layer of either MgO or Al$_2$O$_3$ is much more favorable on a standard Si/SiO$_2$ wafer instead of a graphene surface. However, even in our devices we observe pinholes\cite{drogeler_12ns_2016} and a significant doping of the areas of graphene that are in contact with the electrodes.\cite{drogeler_nanosecond_2014}

Another factor that is impacting the performance of contacts in spintronic applications is the possible contamination of the contact-to-graphene interface with hydrocarbons. If contacts are fabricated on top of graphene, the graphene is unavoidably exposed to the resist and every chemical that is needed for the lithography process, which leads to a contamination of graphene.\cite{Lin-GrapheneAnnealingHow-2012} But even in our fabrication method the contact-to-graphene interface is very likely contaminated by hydrocarbons. On the one hand, resist residues can be deposited on top of the tunnel barrier during the lift-off process. On the other hand, during the time between exfoliation and the transfer of the graphene flake on top of the electrodes, the graphene flake will be slowly but steadily covered with airborne hydrocarbons.\cite{Li-EffectAirborneContaminants-2013,Martinez-ContaminantsGraphitic-2013} It is an ongoing investigation from our side, if variations in the pressure that is applied during the transfer process might squeeze out hydrocarbon contamination at the contact-to-graphene interface to different extents, or if the highly varying contact resistances between different devices are only due to randomly distributed pinholes in the tunnel barrier.

All in all, it is extremely challenging to fabricate high-quality, non-invasive contacts to graphene in a reproducible manner.
Therefore, we are convinced that once the graphene flakes are of sufficiently high quality, any variation from device to device will be caused by variations in contact properties rather than variations in graphene quality. In accordance to this assumption, we e.g.\,do not observe a clear correlation between the measured spin lifetimes and aspects like 1.)\:charge carrier mobility, 2.)\:number of graphene layers (single layer, bilayer, or even trilayer graphene), or 3.)\:if the graphene between the contacts is suspended or not.\cite{drogeler_nanosecond_2014,drogeler_nanosecond_2015,drogeler_12ns_2016,drogeler_dry-transferred_2017}

\begin{figure*}[tb]
\includegraphics{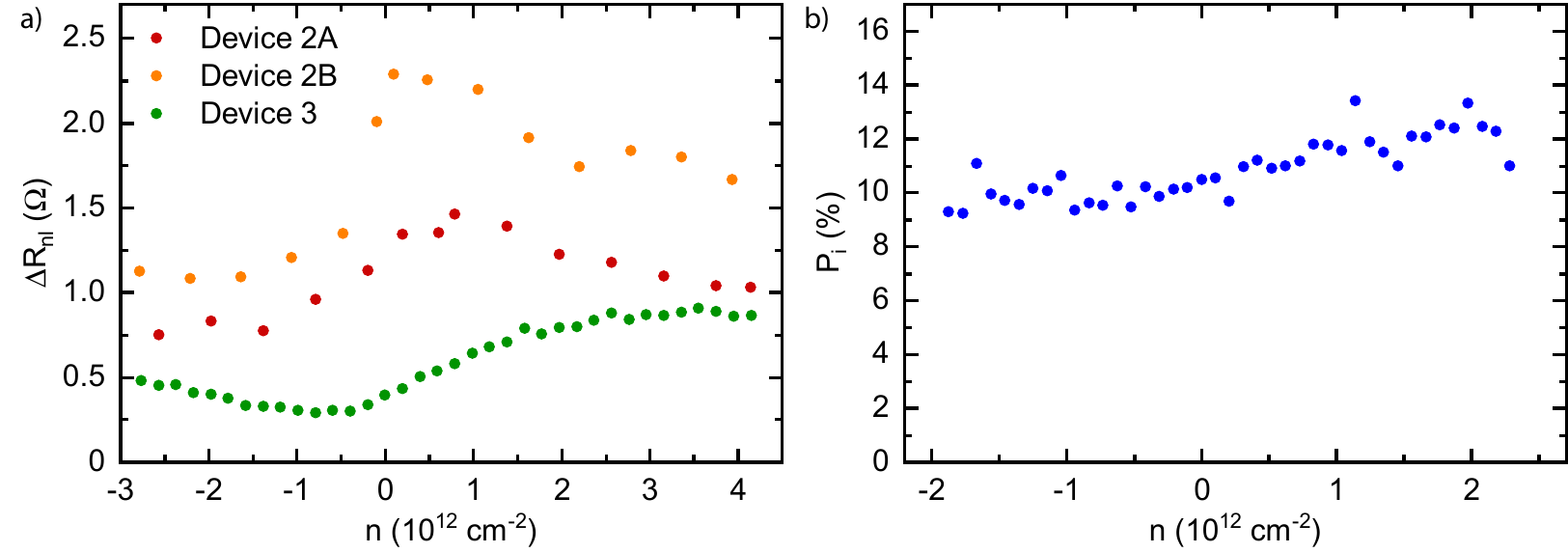} 
\caption{a) Non-local resistance $R_\mathrm{nl}$ of devices 2 and 3 in dependence of the charge carrier density $n$. b) Spin injection efficiency $P_\text{mean}$ of device 1 versus charge carrier density. Notably the values are lower in the hole conduction regime for both $R_\mathrm{nl}$ and $P_i$, i.e.\,for negative $n$.}
\label{fig:SI_Pi_Rnl}
\end{figure*}

\section{Amplitude of the non-local spin signal and injection efficiency}
The gate-dependent amplitude of the non-local spin signal $\Delta R_\mathrm{nl}$, which was determined by the difference between parallel and anti-parallel configuration in the Hanle spin precession measurements at $B=\SI{0}{T}$, is plotted in Fig.\,\ref{fig:SI_Pi_Rnl}a for devices 2A, 2B and 3. Interestingly, we observe a common feature in all of our devices (including device 1, see Fig.\,2c in the main manuscript): The amplitude of the spin signal is always higher in the electron regime ($n>0$) compared to the hole regime ($n<0$). This fact is even more intriguing as the spin lifetime shows a completely different gate dependency, especially for device 1 (see Fig.\,3 in the main manuscript).

Traditionally, the relationship between spin lifetime and spin amplitude is given as:\cite{Tombros2007,Popinciuc2009}
\begin{equation}
\Delta R_\mathrm{nl} = \frac{P_\text{i} P_\text{d} \lambda_\textrm{s}}{W\sigma}\exp{\left(-L/\lambda_\textrm{s}\right)},
\label{efficiency}
\end{equation}
where $P_\text{i}$ is the injection efficiency, $P_\text{d}$ the detection efficiency, $W$ the width of the graphene channel, $\sigma$ the conductivity of graphene, $L$ the distance between injection and detection electrode, and $\lambda_\textrm{s} = \sqrt{D_\textrm{s} \tau_\textrm{s}}$ the spin diffusion length, calculated with the spin diffusion coefficient $D_\textrm{s}$ and the spin lifetime $\tau_\textrm{s}$. According to equation~\ref{efficiency} the increased spin lifetime far from the charge neutrality point in device 1 (see Fig.\,3 in the main manuscript) must be compensated by a less efficient spin injection and/or detection efficiency to explain that the amplitude of the spin signal does not show the same pronounced increase towards larger charge carrier densities.

For comparison to other publication, we calculated the product of the spin injection and detection efficiencies ($P_\text{i} \cdot P_\text{d}$) according to equation~\ref{efficiency} and show its gate-dependent geometric mean $P_\text{mean} = \sqrt{P_\text{i} \cdot P_\text{d}}$ in Fig.\,\ref{fig:SI_Pi_Rnl}b.
We note, however, that this approach is misleading, as in our device design the graphene flake lies on top of the metallic contacts. The contacts therefore screen the electric field from the back gate preventing a tuning of the charge carrier density at the interface between graphene and contacts by a gate voltage. Hence, the spin injection from the contacts into the graphene is not expected to change with the back gate. The same is true for the spin detection.

Instead, the higher spin signal in the electron regime ($n>0$) compared to the hole regime ($n<0$) may be explained by the formation of a p-n-junction between two distinct parts of the graphene flake: The non-tunable part that is lying on top of the contacts and the tunable part between the contacts. We previously have shown by Raman spectroscopy measurements that the graphene on top of the electrodes is n-doped.\cite{drogeler_nanosecond_2014} Therefore, tuning the graphene between the contacts into the hole-regime is creating a p-n-junction somewhere near the edges of the contacts. If this p-n-junction acts as a diffusion barrier for electrons, the residence time of spin polarized charge carriers within the part of graphene that lies on top of the contacts increases. As we believe that the spin lifetime is much lower in these parts compared to the graphene between the contacts (see explanation in previous section), the spins undergo a stronger relaxation. This in turn leads to a reduced "spin injection efficiency" of spins from the graphene part on top of the contacts into the graphene part between the contacts, explaining the reduced amplitude of the spin signal in the hole-regime.

However, further investigations are required to determine the exact impact of such contact-induced p-n-junctions on spin diffusion. Additionally, it has to be investigated if these p-n-junctions or the gate screening of the contact region by the metal electrodes is the reason why the amplitude of the spin signal shows quite a small gate-dependence in all of our samples (see also the data in our previous studies in Refs.\, \citenum{drogeler_nanosecond_2014,drogeler_nanosecond_2015,drogeler_dry-transferred_2017}).

\bibliographystyle{apsrev4-2}